\documentclass[12pt]{article}
\usepackage{amssymb,amsmath,latexsym,graphicx, bm, mathrsfs, hyperref}

\newtheorem{axiom}{Definition}[section]

\newcommand{\be}{\begin{equation}}
\newcommand{\ee}{\end{equation}}

\begin{document}

\title{Jacobi-Maupertuis-Eisenhart metric and geodesic flows  \author{ Sumanto Chanda$^1$, G.W. Gibbons$^2$, Partha Guha$^{1,3}$ }}

\maketitle
\thispagestyle{empty}

\begin{minipage}{0.3\textwidth}
\begin{flushleft}
\textit{\small $ ^1$ S.N. Bose National Centre for Basic Sciences} \\
\textit{\small JD Block, Sector-3, Salt Lake, Calcutta-700098, INDIA.} \\
\texttt{\small sumanto12@bose.res.in \\  partha@bose.res.in}
\end{flushleft}
\end{minipage}
\begin{minipage}{0.3\textwidth}
\begin{center}
\textit{\small $ ^2$  D.A.M.T.P., \\ University of  Cambridge} \\
\textit{\small Wilberforce Road, \\ Cambridge CB3 0WA,  U.K.} \\
\texttt{\small G.W.Gibbons@damtp.cam.ac.uk}
\end{center}
\end{minipage}
\begin{minipage}{0.3\textwidth}
\begin{flushright} \large
\textit{\small $^3$ Institut des Hautes \\ \'Etudes Scientifiques} \\ 
\textit{\small 35 Route de Chartres 91440, \\ Bures-sur-Yvette France} \\ 
\texttt{\small guha@ihes.fr}
\end{flushright}
\end{minipage}

\bigskip

\bigskip


\bigskip

{\bf{Keywords and keyphrases :}} Jacobi-Maupertuis metric, Randers type Finsler metric,\\ geodesics,
Eisenhart-Duval lift.

\smallskip

\abstract{The Jacobi metric derived from the line element by one of the authors 
is shown to reduce to the 
standard formulation in the non-relativistic approximation. 
We obtain the Jacobi metric for various stationary metrics. Finally, the 
Jacobi-Maupertuis metric is formulated for time-dependent metrics by including the Eisenhart-Duval lift, known as the 
Jacobi-Eisenhart metric.}

\tableofcontents

\setcounter{page}{1}

\numberwithin{equation}{section}

\section{Introduction}

Riemann studied concepts like curvature and geodesics by introducing Riemannian manifolds 
in his Habilitationsthesis, where he defined an inner product on every tangent space of 
a manifold. Such inner products were defined via a structure known as the metric that defines 
infinitesimal length elements locally on the tangent space, which can be integrated to compute 
a given path's length \cite{br, ilgm} between any two points on the manifold. The shortest path in 
terms of integrated path length is defined as the geodesic, which according to Maupertuis, is 
effectively the path of least action, comparable to Fermat's path of least time for light \cite{maup}. On 
these manifolds, the form of the action integrals along geodesics is known as the Maupertuis 
form of action \cite{nom} along geodesics, about which the integrand is an exact differential. In this article 
we will focus on geodesics and their projection onto the constant energy hypersurface. \\

The Jacobi metric formulation is a procedure for producing a geodesic from a given hamiltonian. 
Such trajectories of a Hamiltonian system can be viewed as geodesics of a corresponding 
configuration space or its enlargement under some constraints. Since we  parametrize with 
respect to time $\tau = t$, the term quadratic in time is present as the potential. Since such 
Hamiltonians often already arise from Lagrangians originating from a metric, the Jacobi metric formulation 
obtains a lower dimensional geodesic from a higher dimensional one. 

One interesting feature of the Jacobi metric is the effect of its curvature. This was shown 
by Ong in his application of the metric to gravity in \cite{ocp}, where he studied its curvature for 
the Newtonian $n$-body problem (also in \cite{gs}), which for $n = 2$ reduces to the Kepler problem. 
In such cases, where the metric spatial components are flat, the resulting Jacobi-metric is 
conformally flat, which makes evaluating its curvature a simple matter. \\

In this article, we will explore three different, but equivalent approaches to obtaining the 
Jacobi metric. In the first one, we start with the regular formulation of the action with the 
lagrangian for autonomous mechanical systems. We shall cover two ways of formulating the Jacobi 
metric with this approach: by equating the action to a Lorentz invariant line element integral, 
and by redefining the system from a constant energy hypersurface to a unit momentum 
hypersurface, where the kinetic energy is rescaled by a conformal factor to unity \cite{br, avt}. 
We then proceed to obtain the Jacobi metric purely from the line element integral of Rander's form of stationary metric, 
essentially reproducing the formulation used in \cite{gwg},  while the author in \cite{gwg} employed a  static metric and a  Zermelo form for the
stationary metric \cite{ghww}. In 1941, G. Randers \cite{Randers} introduced a Finsler metric by modifying a Riemannian metric 
$g = g_{ij} \ dx^i \otimes dx^j$ by a linear term $b = b_i(x)dx^i$, the resulting norm on the tangent space is given by
$$F(x,y) = \sqrt{g_{ij}y^iy^j} + b_i(x) y^i, \qquad y = y^i\partial_{x^i} \in T_xM. $$
Randers metrics have received much attention \cite{Robles,Brody} lately because these yield the solutions to Zermelo's problem of navigation, most recently, it has been extended to quantum navigation problem of finding the time-optimal control Hamiltonian 
\cite{Brody1}. In \cite{Zermelo}, E. Zermelo studied a classical control problem to find a deviation of geodesics under the 
action of a time-dependent vector field.

\smallskip

In this article, we will go a step further, 
and apply the non-relativistic approximation to this result, thereby reproducing the 
previous result and equating the two formulations. \\ \\
We shall next obtain the Jacobi metric for time-dependent mechanical systems. So far, the 
formulation has only been applied to  time-independent static and stationary metrics. The difficulty 
in application to time-dependent metrics is the absence of a constant energy hypersurface. To 
resolve this issue, we modify the metric via the Eisenhart-Duval lift introduced 
by Eisenhart \cite{eisenhart} and rediscovered by Duval et. al. \cite{dbkp}. This means introducing an extra dimension via a dummy variable and 
a fixed hypersurface on which to project the geodesic, thus relating $n$ dimensional mechanics to 
geodesics on $n+2$ dimensional space. First, we demonstrate the utility of the 
Eisenhart-Duval lift in this context, by describing autonomous and non-autonomous systems 
with and without the lift applied, then deduce the formulation from the line-element integral, 
and finally apply limits for a non-relativistic approximation. We propose calling the result the Jacobi-Eisenhart metric. Finally, we obtain the same results using projective transformations and compare them to verify consistency of the results.

\section{Basic formulation}

Let $g$ be a Riemannian metric on the manifold $M$. If $\dot{x} \in T_x M$, then its length is
$$|| \dot{x} || := \sqrt{g_x (\dot{x}, \dot{x})}.$$
If $\gamma : [a, b] \rightarrow M$ is a smooth curve in $M$, then $\dfrac{d \gamma}{d \tau} \in T_{\gamma(\tau)} M$, which lets us define the length of the curve $\gamma$ \cite{br, ilgm} as
$$l(\gamma) := \int_a^b \bigg| \bigg| \frac{d \gamma (\tau)}{d \tau} \bigg| \bigg|_{\gamma(t)} d \tau.$$
where for the geodesic, the following condition holds:
$$\delta \ l(\gamma) = 0.$$
The geodesic can also be defined as follows:
\begin{axiom}
A geodesic in a pseudo-Riemannian manifold $(M,g)$ is a solution to the 
Euler-Lagrange equations
\begin{equation}
\label{eulag} \big[ \mathcal{L} \big]^x := \frac{d \ }{d t} \bigg( \frac{\partial \mathcal{L}}{\partial \dot{x}^i} \bigg) - \frac{\partial \mathcal{L}}{\partial x^i} = 0.
\end{equation}
where the Lagrangian $\mathcal{L} : TM \rightarrow \mathbb{R}$ is defined by $\mathcal{L} = \frac12 g_x (\dot{x}; \dot{x})$.
\end{axiom}
Consider an integral $I_{12}$ along a path parametrized by $\tau$ between any two points defined by 
\begin{equation}
\label{lint} I_{12} = \int_1^2 d I = \int_1^2 d \tau \ L.
\end{equation}
where $L = L (\bm{x}, \dot{\bm{x}})$ is a function quadratic in velocity, the position being $\bm{x}$ and the velocity being   $\dot{\bm{x}}$. The geodesic is characterised by the Euler-Lagrange equation (\ref{eulag}) which is derivable from
$$\delta I_{12} = \int_1^2 \delta (d I) =  0$$
This means that if we vary the line integral (\ref{lint}) and apply (\ref{eulag}), we have
\[ \begin{split}
\delta I_{12} &= \int_1^2 d \tau \ \delta L (\bm{x}, \dot{\bm{x}}) = \int_1^2 d \tau \bigg( \frac{\partial L}{\partial x^i} \delta x^i + \frac{\partial L}{\partial \dot{x}^i} \delta \dot{x}^i \bigg), \\
&= \int_1^2 d \tau \bigg[ \frac{d \ }{d \tau} \bigg( \frac{\partial L}{\partial \dot{x}^i} \bigg) \delta x^i + \frac{\partial L}{\partial \dot{x}^i} \frac{d \ }{d \tau} \big( \delta x^i \big) \bigg] = \int_1^2 d \tau \frac{d \ }{d \tau} \bigg( \frac{\partial L}{\partial \dot{x}^i} \delta x^i \bigg), \\
& \quad \Rightarrow \quad \int_1^2 \delta (d I) = \int_1^2 d \bigg( \frac{\partial L}{\partial \dot{x}^i} \delta x^i \bigg).
\end{split} \]
Since we are considering the path of extremal variation, we are dealing with an integral that is 
locally exact about the geodesic, (ie. $\delta (d I) = d (\delta I)$). This means on substituting the momentum $p_i = \dfrac{\partial L}{\partial \dot{x}^i}$, the effective integral along the geodesic and the effective lagrangian $L_{geod}$ along the geodesic are given by
$$\int_1^2 d (\delta I) = \int_1^2 d \bigg( \frac{\partial L}{\partial \dot{x}^i} \delta x^i \bigg) \qquad \Rightarrow \qquad \delta I = \frac{\partial L}{\partial \dot{x}^i} \delta x^i. $$
\begin{equation} \label{maup} 
\begin{split}
I_{12} = \int_1^2 \frac{\partial L}{\partial \dot{x}^i} &d x^i = \int_1^2 d \tau \bigg( \frac{\partial L}{\partial \dot{x}^i} \dot{x}^i \bigg) = \int_1^2 p_i d x^i, \\
&L_{geod} = \frac{\partial L}{\partial \dot{x}^i} \dot{x}^i = p_i \dot{x}^i.
\end{split}
\end{equation}
where the effective line integral is known as the Maupertuis form \cite{nom} of the line integral.

\numberwithin{equation}{subsection}

\subsection{Natural Hamiltonian}

If one starts with a static metric ($g_{0i} = 0$) on a given $n+1$ dimensional space-time 
$$dl^2 = g_{\mu \nu} dx^\mu dx^\nu = g_{00} \ dt^2 + g_{ij} \ dx^i dx^j.$$ 
it is a simple matter to formulate the corresponding lagrangian describing the dynamics on 
that space. Such dynamical systems under affine parametrization $\tau = x^0 = t$ are defined by 
the mechanical action and its related lagrangian:
\begin{align}
\label{act} &S = \int_{\tau_1}^{\tau_2} d \tau \ L (\bm{x}, \bm{\dot{x}}). \\
\label{lag} L (\bm{x}, \bm{\dot{x}}) &= \frac m2 g_{\mu \nu} (\bm{x}) \dot{x}^\mu \dot{x}^\nu = \frac m2 g_{ij} \dot{x}^i \dot{x}^j - U (\bm{x}) \equiv T - U (\bm{x}).
\end{align}
If the lagrangian can have a natural form given by (\ref{lag}), then so will the Hamiltonian when momentum has been solved for velocity and substitute back inthe hamiltonian.
The natural hamiltonian for a time-independent dynamical system that acts as the generator 
for time-translations is a conserved quantity is given by a Legendre transformation
$$H (\bm{x}, \bm{p}) = \sum_{i=1}^n p_i \dot{x}^i - L (\bm{x}, \bm{\dot{x}}) \hspace{2cm} p_i = \frac{\partial L}{\partial x^i} = g_{ij} (\bm{x}) \dot{x}^j.$$
\begin{equation}
\label{natham} H (\bm{x}, \bm{p}) = \frac1{2m} g^{ij} (\bm{x}) p_i p_j + U (\bm{x}) \equiv T (\bm{x}, \bm{\dot{x}}) + U (\bm{x}) = E.
\end{equation}
where the dynamical equations or Hamilton's equations of motion are:
\begin{equation}
\label{hamdyn} \begin{split}
\dot{x}^i &= \frac{\partial H}{\partial p_i} = \frac{g^{ij} (\bm{x})}m p_j, \\ 
\dot{p}_i &= \frac{\partial H}{\partial x^i} = \frac1{2m} \frac{\partial g^{ij} (\bm{x})}{\partial x^i} p_i p_j + \frac{\partial U}{\partial x^i}.
\end{split}
\end{equation}
This means that the lagrangian of (\ref{lag}) can be written as
$$L = 2T - E.$$
and thus the zero-variation equation of the action (\ref{act}) can be written as
$$\delta S = \delta \Big( \int_{\tau_1}^{\tau^2} d \tau \ L \Big) = \delta \Big[ \int_{\tau_1}^{\tau^2} d \tau \big( 2T - E \big) \Big] = 2 \int_{\tau_1}^{\tau^2} d \tau \ \delta T.$$
Thus, the effective action is given as:
\begin{equation}
\label{seff} S_{eff} = \int_{\tau_1}^{\tau_2} d \tau \ 2 T.
\end{equation}
Being the generator of time translations, the time derivative of any functions is given by Poisson 
Bracket operations $\dot{f} = \big\{ f, H \big\}$. Naturally, any conserved quantities will be in involution with 
this Hamiltonian, itself being a conserved quantity:
$$\dot{Q} = \big\{ Q, H \big\} = 0.$$
This hamiltonian is made up of 2 parts; quadratic and potential. In the next section, we shall 
see how to reduce it to being homogeneously quadratic.

\subsection{Jacobi metric}

From (\ref{seff}), we can see that for conserved quantities, an alternative formula for the action would 
suffice to describe geodesics with conserved energies. This effective lagrangian based action integral 
may be equated to a metric line-element integral \cite{nom} using (\ref{natham}) as follows:
$$S_{eff} = \int_{\tau_1}^{\tau_2} d \tau \ 2 T = \int_{\tau_1}^{\tau_2} d \tau \ \sqrt{2 T} \sqrt{2 T} = \int_{\tau_1}^{\tau_2} d \tau \ \sqrt{2 (E - U)} \sqrt{m g_{ij} (\bm{x}) \dot{x}^i \dot{x}^j},$$
$$S_{eff} = \int_{\tau_1}^{\tau_2} d \tau \ \sqrt{2m (E - U) g_{ij} (\bm{x}) \dot{x}^i \dot{x}^j} \equiv \int_1^2 d \tau \sqrt{\bigg(\frac{d l_{eff}}{d \tau} \bigg)^2}.$$
Thus, the effective Jacobi metric can be given as
$$d l_{eff}^2 = L dt^2 = 4 \big[ E - U (\bm{x}) \big] T dt^2 \qquad T = \frac m2 g_{ij} (\bm{x}) \dot{x}^i \dot{x}^j,$$
$$\Rightarrow \qquad d l_{eff}^2 = 2m \big( E - U \big) g_{ij} (\bm{x}) dx^i dx^j.$$
We can view the solution curves of natural mechanical systems as the geodesics of a special 
metric. This process allows us to convert the hamiltonian $n+1$ dimensional system into a 
spatial $n$-dimensional geodesic with a rescaled conserved Hamiltonian:
\begin{equation}
\label{jacham} g^{ij} (\bm{x}) p_i p_j = 2m \big[ E - U (\bm{x}) \big] \qquad \Rightarrow \qquad \widetilde{H} = \frac{g^{ij} (\bm{x})}{2m [E - U (\bm{x})]} p_i p_j = 1.
\end{equation}
We have essentially taken the kinetic energy part of the total conserved energy of the system, 
and rescaled it with a conformal factor that is its inverse into an equivalent constrained system 
with unit momentum sphere. This means that the metric and its inverse transform into the 
Jacobi metric as follows:
\begin{equation} \label{jacmet}
\begin{split}
&\widetilde{g}^{ij} (\bm{x}) p_i p_j = 1, \\
\widetilde{g}^{ij} (\bm{x}) = \frac{g^{ij} (\bm{x})}{2m \big[E - U (\bm{x}) \big]} &\qquad \Rightarrow \qquad \widetilde{g}_{ij} (\bm{x}) = 2m \big[ E - U (\bm{x}) \big] g_{ij} (\bm{x}).
\end{split}
\end{equation}
where the kinetic energy part of the system serves as the conformal factor. We can summarise 
the details with the following theorem.

\newtheorem{theorem}{Theorem}

\begin{theorem}[Jacobi-Maupertuis principle]
Let $T : TM \rightarrow \mathbb{R}$ be a smooth pseudo-Riemannian 
metric and let $U : M \rightarrow \mathbb{R}$ be a smooth potential energy function. Let $t \mapsto x(t), I \rightarrow M$ be 
a curve in $M$ such that $H \big(x(t), \frac{d x(t)}{dt} \big) = E \in \mathbb{R}$ and $U (x(t)) \neq E$ for all $t$. Then the map 
$t \mapsto S(t), I \rightarrow \mathbb{R}$ defined by
$$S(t) = 2 \int_0^t d \tau \big[ E - U (x(t)) \big].$$
is a diffeomorphism onto its image $J$. We denote its inverse by $S \mapsto t(S); J \rightarrow I$. Moreover, 
the curve $t \mapsto x(t)$ in $M$ is a solution to the Euler-Lagrange equation $\big[ T - U \big]^x = 0$ (see \ref{eulag}), 
iff the curve $S \mapsto x(t(S)), J \rightarrow M$ is a geodesic of the ``Jacobi metric"
$$\widetilde{T} = (E - U) T.$$
\end{theorem} 
One may ask why according to (\ref{natham}) we are not substituting $T (\bm{x}, \bm{\dot{x}}) = E - U (\bm{x})$. The reason 
is that doing so would effectively make the metric tensor components velocity dependent
$$\widetilde{g}_{ij} (\bm{x}, \bm{\dot{x}}) = T (\bm{x}, \bm{\dot{x}}) g_{ij} (\bm{x}).$$
like the Finsler metric. Naturally, for a transformed Hamiltonian, the dynamical description 
should also change.

\subsection{Conserved quantities and Clairaut's constant}

Starting with the Hamiltonian in (\ref{jacham}), we shall write the dynamical equations with respect 
to a new parameter $s$ as shown in \cite{br, avt}
\begin{equation}
\label{jacdyn} \begin{split}
\frac{d x^i}{d s} &= \frac{\partial \widetilde{H}}{\partial p_i} = \frac{g^{ij} (\bm{x})}{2m \big[ E - U (\bm{x}) \big]} p_j, \\
\frac{d p_i}{d s} &= - \frac{\partial \widetilde{H}}{\partial x^i} = - \frac1{2m \big[ E - U (\bm{x}) \big]} \bigg[ \frac12 \frac{\partial g^{ij} (\bm{x})}{\partial x^i} p_i p_j + \frac{\partial U}{\partial x^i} \bigg].
\end{split}
\end{equation}
Upon comparison with (\ref{hamdyn}), we can see that the dynamical equations are unaltered, except 
for a reparametrization as in \cite{br, avt}, given by:
\begin{equation}
\label{repara} \frac{d s}{d t} = 2m \big[ E - U (\bm{x}) \big].
\end{equation}
Consequently, for any conserved quantity $K = K^{(2)ij} p_i p_j + K^{(0)}$, we can say:
\begin{align}
\frac{d K}{d s} = \big\{ K, \widetilde{H} \big\} &= \frac{d t}{d s} \frac{d K}{d t} = \frac1{2m \big[ E - U (\bm{x}) \big]} \big\{ K, H \big\}. \\ 
\therefore \qquad \big\{ K, \widetilde{H} \big\} &= 0 \qquad \Rightarrow \qquad \big\{ K, H \big\} = 0.
\end{align}
In \cite{th}, T. Houri describes $\widetilde{K} = K^{(2)ij} p_i p_j + K^{(0)} \widetilde{H}$ where according to (\ref{jacham}), we can say
\begin{equation}
\label{cnsvd} \widetilde{K} = K^{(2)ij} p_i p_j + K^{(0)} \widetilde{H} \quad = \quad K^{(2)ij} p_i p_j + K^{(0)} = K \qquad \qquad \because \quad \widetilde{H} = 1.
\end{equation}
Thus, showing that the conserved quantities remain the same for the Jacobi metric. Taking angular momentum for example, if the spatial metric exhibits spherical symmetry, as described below:
\begin{equation}
\label{sphsym} g_{ij} (\bm{x}) dx^i dx^j = W^2 (\bm{x}) dr^2 + r^2 \big( d \theta^2 + \sin^2 \theta \ d \varphi^2 \big).
\end{equation}
Then using (\ref{repara}), we will have the conserved angular momentum for $\theta = \frac{\pi}2$ in the form known as Clairaut's 
constant given by:
\begin{equation}
\label{angmom} R = 2m r^2 \Big( E - U (\bm{x}) \Big) \frac{d \varphi}{d s} = m r^2 \frac{d \varphi}{d \tau} = const.
\end{equation}
showing that the angular momentum $R$ in (\ref{angmom}), as a first integral is invariant under such formulation as shown in \ref{cnsvd}.

\subsection{Formulation from a metric line element}

One of the authors  formulated the Jacobi metric from the line element in \cite{gwg} and demonstrated the 
formulation for  the Schwarzschild metric. Here, we will show how the line element formulation 
equates to that given by (\ref{jacmet}) which describes the non-relativistic formulation. \\

It is worth  noting  that in  \cite{gwg}, the Jacobi metric was formulated only for static metrics and stationary metrics of the Zermelo form.
Here we formulate the Jacobi metric for stationary metrics of the Randers form of Finsler metric. 
Stationary metrics (with vector potential terms $A_i \neq 0$) are distinct from static metrics in the sense that while both are time-translation invariant, only static metrics are time-reversal invariant. If  $A_i = 0$  stationary metrics reduce to  static metrics. \\ 

Let us consider the following metric:
\begin{equation}
d l^2 = - c^2 V^2 ( \bm{x} ) \big( d t + A_i ( \bm{x} ) dx^i \big)^2 + g_{ij} ( \bm{x} ) dx^i dx^j.
\end{equation}
and the corresponding Lagrangian is given as:
\begin{equation}
L ( \bm{x}, \dot{\bm{x}} ) = m \sqrt{c^2 V^2 ( \bm{x} ) \big( \dot{t} + A_i ( \bm{x} ) \dot{x}^i \big)^2 - g_{ij} ( \bm{x} ) \dot{x}^i \dot{x}^j}.
\end{equation}
The momentum conjugate to co-ordinates are given by:
\begin{equation} \label{relmom}
\begin{split}
\frac Hc = \frac{\partial L}{\partial \dot{t}} &= \frac{m c^2 V^2 ( \bm{x} ) \big( \dot{t} + A_k ( \bm{x} ) \dot{x}^k \big)}{\sqrt{c^2 V^2 ( \bm{x} ) \big( \dot{t} + A_k ( \bm{x} ) \dot{x}^k \big)^2 - g_{ij} ( \bm{x} ) \dot{x}^i \dot{x}^j}} = \frac{\mathcal{E}}c, \\
\frac{p_i}c = \frac{\partial L}{\partial \dot{x}^i} &= \frac{m \big\{ c^2 V^2 ( \bm{x} ) A_i ( \bm{x} ) \big( \dot{t} + A_k ( \bm{x} ) \dot{x}^k \big) - g_{ij} ( \bm{x} ) \dot{x}^j \big\}}{\sqrt{c^2 V^2 ( \bm{x} ) \big( \dot{t} + A_k ( \bm{x} ) \dot{x}^k \big)^2 - g_{ij} ( \bm{x} ) \dot{x}^i \dot{x}^j}}.
\end{split}
\end{equation}
With the following calculations using (\ref{relmom}), we will have
\[ \begin{split}
\bigg( \frac{\mathcal{E}}c \bigg)^2 - m^2 c^2 V^2 ( \bm{x} ) &= m^2 c^2 V^2 ( \bm{x} ) \bigg[ \frac{c^2 V^2 ( \bm{x} ) \big( \dot{t} + A_k ( \bm{x} ) \dot{x}^k \big)^2}{c^2 V^2 ( \bm{x} ) \big( \dot{t} + A_k ( \bm{x} ) \dot{x}^k \big)^2 - g_{ij} ( \bm{x} ) \dot{x}^i \dot{x}^j} - 1 \bigg], \\
&= \frac{m^2 c^2 V^2 ( \bm{x} ) g_{ij} ( \bm{x} ) \dot{x}^i \dot{x}^j}{c^2 V^2 ( \bm{x} ) \big( \dot{t} + A_k ( \bm{x} ) \dot{x}^k \big)^2 - g_{ij} ( \bm{x} ) \dot{x}^i \dot{x}^j}.
\end{split} \]
{ From (\ref{relmom})}, we can see that the gauge-covariant momenta are given by:
$$\frac{\Pi_i}c = \frac{p_i}c - \frac{m c^2 V^2 ( \bm{x} ) A_i ( \bm{x} ) \big( \dot{t} + A_j \dot{x}^j \big)}{\sqrt{c^2 V^2 ( \bm{x} ) \big( \dot{t} + A_k ( \bm{x} ) \dot{x}^k \big)^2 - g_{ij} ( \bm{x} ) \dot{x}^i \dot{x}^j}} = \frac{ - m g_{ij} ( \bm{x} ) \dot{x}^j }{\sqrt{c^2 V^2 ( \bm{x} ) \big( \dot{t} + A_k ( \bm{x} ) \dot{x}^k \big)^2 - g_{ij} ( \bm{x} ) \dot{x}^i \dot{x}^j}},$$
\begin{equation}
\label{tljacham} \mathcal{E}^2 - m^2 c^4 V^2 ( \bm{x} ) = c^2 V^2 ( \bm{x} ) g^{ij} ( \bm{x} )  \Pi_i \Pi_j \qquad \Rightarrow \qquad \frac{c^2 V^2 ( \bm{x} ) g^{ij} ( \bm{x} )}{\mathcal{E}^2 - m^2 c^2 V^2 ( \bm{x} )} \Pi_i \Pi_j = 1.
\end{equation}
One can easily see that in the flat space setting $V^2 ( \bm{x} ) = 1$ in (\ref{tljacham}), we have the familiar relativistic energy equation
$$\mathcal{E}^2 = \big| \Pi \big|^2 c^2 + m^2 c^4.$$
Thus from the inverse metric (\ref{tljacham}) we have the Jacobi metric given by:
\begin{equation}
\label{tljacmet} J^{ij} ( \bm{x} ) = \frac{c^2 V^2 ( \bm{x} ) g^{ij} ( \bm{x} )}{\mathcal{E}^2 - m^2 c^4 V^2 ( \bm{x} )} \qquad \Rightarrow \qquad J_{ij} ( \bm{x} ) = \frac{ \mathcal{E}^2 - m^2 c^4 V^2 ( \bm{x} ) }{c^2 V^2 ( \bm{x} )} g_{ij} ( \bm{x} ).
\end{equation}
Thus, for a fixed relativistic energy $\mathcal{E}$, all timelike geodesics are geodesics of the above Jacobi 
metric. Now that we have summarized the formulation of the Jacobi metric for time-like 
geodesics, we shall see how it evolves under the non-relativistic approximation. Suppose that 
we write the temporal metric component as
\begin{equation}
\label{relpot} V^2 ( \bm{x} ) = 1 + \frac{2 U (\bm{x})}{mc^2}.
\end{equation}
and set the non-relativistic approximation rules
\begin{equation}
\label{nrelapprox} 2 U (\bm{x}) << mc^2 \qquad \qquad g^{ij} (\bm{x}) \Pi_i \Pi_j << m^2 c^2.
\end{equation}
From \ref{tljacham}, we can see that on applying (\ref{relpot}) and (\ref{nrelapprox}), we get
\[ \begin{split}
\mathcal{E} &= mc^2 \sqrt{1 + \frac{2 U (\bm{x})}{mc^2}} \sqrt{ 1 + \frac{g^{ij} ( \bm{x} )  \Pi_i \Pi_j}{m^2 c^2} } \\
&\approx \bigg( 1 + \frac{U (\bm{x})}{mc^2} + . . . \bigg) \bigg( mc^2 + \frac12 \frac{g^{ij} ( \bm{x} )  \Pi_i \Pi_j}m + . . . \bigg) = mc^2 +  \frac12 \frac{g^{ij} ( \bm{x} )  \Pi_i \Pi_j}m + U (\bm{x}) + . . . . ,
\end{split} \]
$$\therefore \qquad \mathcal{E} \approx mc^2 +  \frac12 \frac{g^{ij} ( \bm{x} )  \Pi_i \Pi_j}m + U (\bm{x}) = mc^2 + T + U (\bm{x}).$$
which assures us that our approximation is on the right track. We shall now rewrite the energy 
in the following form:
\begin{equation}
\label{erule} 
\begin{split}
\mathcal{E} \approx mc^2 &+ E \qquad E = T + U (\bm{x}) << mc^2, \\
\bigg(\frac{\mathcal{E}}{mc^2} \bigg)^2 &= \bigg( 1 + \frac{E}{mc^2} \bigg)^2 \approx 1 + \frac{2 E}{mc^2}.
\end{split}
\end{equation}
We will now see that the Jacobi metric as demonstrated in (\ref{tljacmet}) under the approximations 
of (\ref{nrelapprox}) and (\ref{erule}) becomes
\vspace{-0.25cm}
\[ \begin{split}
J_{ij} ( \bm{x} ) &= \frac{ \mathcal{E}^2 - m^2 c^4 V^2 ( \bm{x} ) }{c^2 V^2 ( \bm{x} )} g_{ij} ( \bm{x} ) = \frac{ \bigg( \dfrac {\mathcal{E}}{mc^2} \bigg)^2 - V^2 ( \bm{x} ) }{\bigg( \dfrac{V ( \bm{x} )}{mc} \bigg)^2} g_{ij} ( \bm{x} ), \\
&\approx \frac{ \bigg( 1 + \dfrac{2 E}{mc^2} \bigg) - \bigg( 1 + \dfrac{2 U (\bm{x})}{mc^2} \bigg) }{\dfrac1{(mc)^2} \bigg( 1 + \dfrac{2 U (\bm{x})}{mc^2} \bigg)} g_{ij} ( \bm{x} ) = \frac{ 2 m \big( E - U (\bm{x}) \big)}{\bigg(1 + \dfrac{2 U ( \bm{x} )}{mc^2}\bigg)} g_{ij} ( \bm{x} ), \\
&\approx 2 m \big( E - U (\bm{x}) \big) \bigg(1 - \dfrac{2 U ( \bm{x} )}{mc^2}\bigg) g_{ij} ( \bm{x} ) \approx 2 m \big( E - U (\bm{x}) \big) g_{ij} ( \bm{x} ),
\end{split} \]
\begin{equation}
\therefore \qquad \boxed{J_{ij} ( \bm{x} ) = 2 m \big( E - U (\bm{x}) \big) g_{ij} ( \bm{x} )}.
\end{equation}
Thus, the Jacobi metric in the non-relativistic approximations agrees with the result 
(\ref{jacmet}), showing that both formulations of a projection of the geodesic onto the constant energy 
hypersurface are consistent and correct.

\subsection{Gaussian curvature of conformally flat spaces}

Now we shall compute the Gaussian curvature for conformally flat Jacobi-metric spaces, using 
Jacobi-Kepler spaces as an example. We shall only consider motion in two dimensions because 
of angular momentum conservation in a radial potential. \\ \\
Thus, the Jacobi-metric is given as a conformally flat metric:
\begin{equation}
d \widetilde{l}^2 = \big( E - U(r) \big) \big( d r^2 + r^2 d \theta^2 \big) = f^2 (r) \big( d r^2 + r^2 d \theta^2 \big).
\end{equation}
Here, the Gaussian curvature is given by:
\[ \begin{split}
e^r = f(r) \ dr \qquad &\qquad e^\theta = r f(r) \ d \theta, \\
d e^\theta = \big( r f(r) \big)' dr \wedge d \theta \qquad &\Rightarrow \qquad {\omega^\theta}_r = \frac{\big( r f(r) \big)'}{f(r)} d \theta, \\ \\
d {\omega^\theta}_r = \bigg( \frac{\big( r f(r) \big)'}{f(r)} \bigg)' dr \wedge d \theta \qquad &\Rightarrow \qquad {R^\theta}_{r \theta r} = - \frac1{r f^2(r)} \bigg( \frac{\big( r f(r) \big)'}{f(r)} \bigg)',
\end{split} \]
\begin{equation}
\label{gacurv} \therefore \qquad K_G = {R^\theta}_{r \theta r} = - \frac1{r f^2 (r)} \frac{d \ }{dr} \bigg( \frac1{f(r)} \frac{d \ }{dr} \big( r f(r) \big) \bigg).
\end{equation}
Thus, for $f^2 (r) = E - U(r)$, the Gaussian curvature (\ref{gacurv}) in this case is given as:
\begin{equation} 
K_G = \frac{\big( r U' (r) \big)' \big( E - U(r) \big) + r \big( U'(r) \big)^2 }{2 r \big( E - U(r) \big)^3}.
\end{equation}
If $h$ is a regular value of $U(r)$ on the boundary ring, ie. $U(r) = h; x \in \partial M$ we have by 
continuity
\begin{equation}
\big( r U' (r) \big)' \big( E - U(r) \big) + r \big( U'(r) \big)^2 > 0, \qquad K_G \longrightarrow \infty.
\end{equation}
In case of the Kepler problem, we have $U(r) = - \dfrac kr$, so the Gaussian curve $K_G$ is:
\begin{equation}
\label{kcurv} K_G = - \frac{k E}{2 \big( r E + k \big)^3}.
\end{equation}
Thus, we can see that the curvature is classified as:
\begin{equation}
 \forall \quad E > - \frac kr \qquad
\begin{cases}
E < 0 \qquad \Rightarrow \qquad K_G > 0 \quad ; \qquad \text{ellipse} \\
E = 0 \qquad \Rightarrow \qquad K_G = 0 \quad ; \qquad \text{parabola} \\
E > 0 \qquad \Rightarrow \qquad K_G < 0 \quad ; \qquad \text{hyperbola} 
\end{cases}.
\end{equation}
Thus, for the Kepler problem, for negative energies in the range $- \dfrac kr < E < 0$, we will have 
posetive curvature, and thus closed periodic orbits described by the Jacobi-Kepler metric. 
What motivates us to connect this theory with the Kepler problem is that it describes $\widetilde{H} = 1$ 
geodesic flow on $T^* S^3, K_G = 1$ energy surface.

\subsection{Schwarzschild metric}

Now that we have summarized the formulation of the Jacobi metric for time-like geodesics, we 
shall demonstrate the third author's  application for the formulation on the Schwarzschild metric \cite{gwg}. 
For the Schwarzschild metric (setting $c = 1$) we are dealing with the case where $A_i (\bm{x}) = 0$ given by:
\begin{equation}
d l^2 = - \bigg( 1 - \frac{2 M}r \bigg) d t^2 + \bigg( 1 - \frac{2 M}r \bigg)^{-1} d r^2 + r^2 \big( d \theta^2 + \sin^2 \theta \ d \varphi^2 \big).
\end{equation}
We can say that
\begin{equation}
V^2 ( \bm{x} ) = \bigg( 1 - \frac{2 M}r \bigg) \hspace{1cm} g_{ij} dx^i dx^j = \bigg( 1 - \frac{2 M}r \bigg)^{-1} d r^2 + r^2 \big( d \theta^2 + \sin^2 \theta \ d \varphi^2 \big).
\end{equation}
Thus, the relativstic Schwarzschild Jacobi metric according to (\ref{tljacmet}) is given by
\begin{equation}
J_{ij} ( \bm{x} ) dx^i dx^j = \bigg[ \mathcal{E}^2 - m^2 \bigg( 1 - \frac{2 M}r \bigg) \bigg] \bigg[ \bigg( 1 - \frac{2 M}r \bigg)^{-2} d r^2 + \bigg( 1 - \frac{2 M}r \bigg)^{-1} r^2 \big( d \theta^2 + \sin^2 \theta \ d \varphi^2 \big) \bigg].
\end{equation}
and the non-relativistic Schwarzschild Jacobi metric according to (\ref{jacmet}) is given by
\begin{equation}
\widetilde{g}_{ij} ( \bm{x} ) dx^i dx^j = 2m \bigg[ E + \frac{m M}r \bigg] \bigg[ \bigg( 1 - \frac{2 M}r \bigg)^{-1} d r^2 + r^2 \big( d \theta^2 + \sin^2 \theta \ d \varphi^2 \big) \bigg].
\end{equation}
Now we shall formulate the Jacobi metric for other geodesics in  {\it stationary spacetimes}.

\section{Jacobi metric for time-like geodesics in stationary spacetime}

Here, we shall apply the present formulation of the Jacobi-metric by the  to other static  and 
{\it stationary} space-time metrics such as Taub-NUT, Bertrand and Kerr metrics.

\subsection{The Taub-NUT metric}

In 1951, Abraham Huskel Taub found an exact solution of Einstein's equations, which was 
subsequently extended to a larger manifold by E. Newman, T. Unti and L. Tamburino in 
1963, known as the the Taub-NUT \cite{extn}. It is a gravitational anti-instanton with corresponding 
$\rm SU(2)$ gauge fields, with geodesics which approximately describe the motion of well seperated 
monopole-monopole interactions. As a dynamical system it exhibits spherically symmetry, 
with geodesics admitting Kepler-type symmetry. \\ \\
The Euclidean Taub-NUT metric is given by:
\begin{equation}
d l^2 = 4M^2 \frac{r - M}{r + M} \big( d \psi + \cos \theta \ d \varphi \big)^2 + \frac{r + M}{r - M} dr^2 + \big( r^2 - M^2 \big) \big( d \theta^2 + \sin^2 \theta \ d \varphi^2 \big).
\end{equation}
where $\psi \equiv t$. However, it is not a spacetime due to the Euclidean signature, which results 
in a slightly different form of Jacobi metric derived by the same approach. Furthermore, 
the nature of its potential term distinguishes it from other spacetimes, such that the lower 
energy and weak potential limits (for other spacetimes we shall see that $V^2 ( \bm{x} )_{M = 0} = 1$) need to be differently 
defined. Here, we can see that
\begin{equation}
\begin{split}
V^2 ( \bm{x} ) &= 4M^2 \frac{r - M}{r + M} \qquad \qquad V^2 ( \bm{x} )_{M = 0} = 0, \\
g_{ij} dx^i dx^j &= \frac{r + M}{r - M} dr^2 + \big( r^2 - M^2 \big) \big( d \theta^2 + \sin^2 \theta \ d \varphi^2 \big).
\end{split}
\end{equation}
Thus, the geometric line-element based Jacobi metric derived in the same manner as (\ref{tljacmet}) 
is given by
\begin{equation}
J_{ij} ( \bm{x} ) dx^i dx^j = \frac{\big( r + M \big)^2 }{4 M^2}\bigg( 4m^2 M^2 \frac{r - M}{r + M} - \mathcal{Q}^2 \bigg) \bigg[ \frac{dr^2}{\big( r - M \big)^2} + \big( d \theta^2 + \sin^2 \theta \ d \varphi^2 \big) \bigg].
\end{equation}
where $\mathcal{Q} = m \dfrac{\partial \ }{\partial \dot{\psi}} \sqrt{\bigg( \dfrac{d l}{d \tau} \bigg)^2}$. On the other hand, the lagrangian based Jacobi metric derived in 
the same manner as (\ref{jacmet}) (according to (\ref{maup}), $E = \sum_\mu p_\mu \dot{x}^\mu - L_{geod} = 0$) is given by
\begin{equation}
\widetilde{g}_{ij} ( \bm{x} ) dx^i dx^j = - Q^2 \frac{\big( r + M \big)^2}{4M^2} \bigg[ \frac{dr^2}{\big( r - M \big)^2} + \big( d \theta^2 + \sin^2 \theta \ d \varphi^2 \big) \bigg].
\end{equation}
which describes the weak potential limit $V^2 ( \bm{x} ) \approx 0$, where $Q = \dfrac m2 \dfrac{\partial \ }{\partial \dot{\psi}} \bigg[ \bigg( \dfrac{d l}{d \tau} \bigg)^2 \bigg]$ is a conserved 
quantity. Now we shall turn our attention to another case: the Bertrand spacetime metric.

\subsection{The Bertrand spacetime metric}

According to Bertrand's theorem, all bounded, closed and periodic orbits in Euclidean space 
are associated only with two potentials: the Kepler-Coloumb $U(r) = \frac ar + b$ and the Hooke-
Oscillator $U(r) = ar^2 + b$, which are dual to each other, related via the Bohlin-Arnold-
Vasiliev transformation \cite{dual, TNdynamics}. The Taub-NUT metric previously discussed is effectively a 
Euclidean Bertrand spacetime metric with magnetic fields applied and exhibits the same 
duality as shown in \cite{TNdynamics}. Perlick showed that Bertrand's theorem arises in General Relativity as well \cite{vp}. 
The Bertrand spacetime metric is given as:
\begin{equation}
d l^2 = - \frac{d t^2}{\Gamma (r)} + h^2 (r) dr^2 + r^2 \big( d \theta^2 + \sin^2 \theta \ d \varphi^2 \big).
\end{equation}
Since angular momentum is conserved under spherical symmetry, taking $\theta = \frac{\pi}2$ and defining $\frac1{\Gamma (r)} = 1 + \frac{2 U (r)}m$, the natural Hamiltonian is given as:
\begin{equation}
H (\bm{x}, \bm{p}) = \frac{p_r^2}{2 h^2 (r)} + \frac{p_\varphi^2}{2 r^2} + \frac m2\bigg( \frac1{\Gamma (r)} - 1 \bigg) = E.
\end{equation}
Therefore, the Hamilton's dynamical equations are:
\begin{equation}
\begin{split}
\dot{r} = \frac{\partial H}{\partial p_r} &= \frac{p_r}{h^2 (r)} \hspace{1.5cm}
\dot{p}_r = - \frac{\partial H}{\partial r} = \frac{p_r^2}{h^2 (r)} \frac{h' (r)}{h (r)} + \frac{p_\varphi^2}{r^3} + \frac{m \Gamma'(r)}{2 \Gamma^2 (r)}, \\
\therefore \hspace{1cm} \dot{p}_r &= \bigg( 2 E  + m- \frac m{\Gamma (r)} \bigg) \frac{h' (r)}{h (r)} + \bigg( \frac1r - \frac{h' (r)}{h (r)} \bigg) \frac{p_\varphi^2}{r^2} + \frac{m \Gamma'(r)}{2 \Gamma^2 (r)}.
\end{split}
\end{equation}
The radial equation of motion is:
$$\ddot{r} = - \bigg( 2 E + m - \frac1{\Gamma (r)} \bigg) \frac{h' (r)}{h^3 (r)} + \bigg( \frac1r + \frac{h' (r)}{h (r)} \bigg) \frac{p_\varphi^2}{h^2 (r) r^2} + \frac{m \Gamma'(r)}{2 h^2 (r) \Gamma^2 (r)}.$$
which for the Kepler problem $U (r) = - \frac kr, h^2 (r) = 1$ is:
$$\ddot{r} = \frac{p_\varphi^2}{r^3} - \frac k{r^2}.$$
By regular formulation, the Jacobi metric is given as:
\begin{equation}
\widetilde{g}_{ij} (\bm{x}) dx^i dx^j = \bigg[ E + \frac m2 \bigg( 1 - \frac1{\Gamma (r)} \bigg) \bigg] \big[ h^2 (r) dr^2 + r^2 \big( d \theta^2 + \sin^2 \theta \ d \varphi^2 \big) \big].
\end{equation}
for which the reparametrized Hamilton's equations according to \ref{jacdyn} are:
\begin{align}
\frac{d r}{d s} &= \frac{d t}{d s} \dot{r} = \frac{2 \Gamma (r)}{(2 E + m) \Gamma (r) - m} \frac{p_r}{h^2 (r)}. \\
\frac{d p_r}{d s} &= \frac{d t}{d s} \dot{p}_r = \frac{2 \Gamma (r)}{(2 E + m) \Gamma (r) - m} \bigg[ \bigg( 2 E  + m- \frac m{\Gamma (r)} \bigg) \frac{h' (r)}{h (r)} + \bigg( \frac1r - \frac{h' (r)}{h (r)} \bigg) \frac{p_\varphi^2}{r^2} + \frac{m \Gamma'(r)}{2 \Gamma^2 (r)} \bigg].
\end{align}
For example, if we consider the Kepler problem, we set $U (r) = - \frac kr, h^2 (r) = 1$ and we have:
\begin{align}
\frac{d r}{d s} &= \frac{d t}{d s} \dot{r} = \frac{2 r}{2 E r + k} p_r. \\
\frac{d p_r}{d s} &= \frac{d t}{d s} \dot{p}_r = \frac{2 r}{2 E r + k} \bigg( \frac{p_\varphi^2}{r^3} - \frac k{r^2} \bigg).
\end{align}
However, if we were to apply the treatment for time-like geodesics of \cite{gwg} where $c^2 V^2 (r) = \frac1{\Gamma (r)}$, then we would have the time-like Jacobi Bertrand metric as per \ref{tljacmet} is
\begin{equation}
J_{ij} (\bm{x}) dx^i dx^j = \bigg( \mathcal{E}^2 \Gamma (r) - m^2 c^2 \bigg) \big[ h^2 (r) dr^2 + r^2 \big( d \theta^2 + \sin^2 \theta \ d \varphi^2 \big) \big].
\end{equation}
The next metric we shall deal with is the Kerr metric.

\subsection{The Kerr metric}

In \cite{dsg}, the Jacobi metric of the Reissner-N\"ordstrom spacetime was given. Here, we shall 
turn our attention to another black-hole spacetime known as the rotating (Kerr) black hole. This is a stationary metric.   \\ \\
The Kerr metric (setting $c = 1$) is:
\begin{equation}
\begin{split}
d l^2 = - \bigg( 1 - \frac{2 G M r}{\rho^2} \bigg) dt^2 &- \frac{4 G M a r \sin^2 \theta}{\rho^2} d \phi \ d t \\
& \quad + \frac{\rho^2}{\Delta} dr^2 + \rho^2 \ d \theta^2 + \frac{\sin^2 \theta}{\rho^2} \bigg[ \big( r^2 + a^2 \big)^2 - a^2 \Delta \sin^2 \theta \bigg] d \phi^2, \\ \\
\Delta (r) = r^2 &- 2 G M r + a^2 \qquad \rho^2 (r, \theta) = r^2 + a^2 \cos^2 \theta.
\end{split}
\end{equation}
Here, the potential term $V^2 (\bm{x})$ and the the spatial metric $g_{ij} (\bm{x})$ are taken to be
\begin{align}
\label{kerrpot} V^2 (\bm{x}) &= 1 - \frac{2 G M r}{\rho^2}. \\
\label{kerrspmet} g_{ij} (\bm{x}) &= \frac{\rho^2}{\Delta} dr^2 + \rho^2 \ d \theta^2 + \frac{\sin^2 \theta}{\rho^2} \bigg[ \big( r^2 + a^2 \big)^2 - a^2 \Delta \sin^2 \theta \bigg] d \phi^2.
\end{align}
So, using the potential (\ref{kerrpot}) and the relativistic Jacobi metric formulation \ref{tljacmet} gives us
\begin{equation}
J_{ij} ( \bm{x} ) dx^i dx^j = \bigg( \frac{\mathcal{E}^2 \rho^2}{\rho^2 - 2 G M r} - m^2 \bigg) \bigg[ \frac{\rho^2}{\Delta} dr^2 + \rho^2 \ d \theta^2 + \frac{\sin^2 \theta}{\rho^2} \bigg\{ \big( r^2 + a^2 \big)^2 - a^2 \Delta \sin^2 \theta \bigg\} d \phi^2 \bigg].
\end{equation}
while the non-relativistic Jacobi metric formulation \ref{jacmet} gives us
\begin{equation}
\widetilde{g}_{ij} ( \bm{x} ) dx^i dx^j = \bigg( E +\frac{2GMr}{\rho^2} \bigg) \bigg[ \frac{\rho^2}{\Delta} dr^2 + \rho^2 \ d \theta^2 + \frac{\sin^2 \theta}{\rho^2} \bigg\{ \big( r^2 + a^2 \big)^2 - a^2 \Delta \sin^2 \theta \bigg\} d \phi^2 \bigg].
\end{equation}
Now we shall consider how to execute such a formulation for time-dependent systems.

\numberwithin{equation}{section}

\section{Jacobi metric for time-dependent systems}

Time dependent systems are essentially those where we find that the system energy is not conserved. 
Usually such systems are dissipative in nature. When we formulate the Jacobi metric 
for autonomous or time-independent systems, we are essentially projecting the geodesic to a 
constant energy hypersurface. However, a non-autonomous or time-dependent system does 
not possess a fixed energy hypersurface, requiring us to improvise our approach. One way to 
deal with time-dependent systems is the Eisenhart-Duval lift. 

The Eisenhart-Duval lift, developed by L.P. Eisenhart \cite{eisenhart} and rediscovered by C. Duval 
\cite{dbkp}, with applications demonstrated in \cite{ca, cghhz} embeds non-relativistic theories into Lorentzian 
geometry. It is one example of a method for geometrizing interactions, where a classical system 
in $n$ dimensions is shown to be dynamically equal to a Lorentzian $n+2$ spacetime. It provides 
a relativistic framework to study nonrelativistic physics, simplifying the study of symmetries of 
a Hamiltonian system by looking at geodesic Hamiltonians. The hidden symmetries of this lift were studied from the perspective of the Dirac equation by Cariglia \cite{cariglia}, and it was applied to study the 
projective and conformal symmetries and quantisation of dissipative systems such as Caldirola 
and Kannai’s damped simple harmonic oscillator in \cite{cdgh}. \\ \\
Let $(M, g)$ be a pseudo-Riemannian manifold, ie. $g$ is a non-degenerate symmetric two times 
covariant tensor field on $M$. Given a local chart $(U, x^1, . . . . x^n)$ on $M$, the local expression for 
$g$ is is given by:
$$g = g_{ij} (\bm{x}) dx^i \otimes dx^j.$$
and the corresponding metric is
\begin{equation}
d l^2 = g_{ij} (\bm{x}) dx^i dx^j.
\end{equation}
The geodesic of the equation is
$$\ddot{x}^i + \Gamma^i_{jk} \dot{x}^j \dot{x}^k = 0.$$
where the connection
$$\Gamma^i_{jk} = \frac12 g^{il} (\bm{x}) \bigg( \frac{\partial g_{lj}}{\partial x^k} + \frac{\partial g_{lk}}{\partial x^j} - \frac{\partial g_{jk}}{\partial x^l} \bigg).$$
can be obtained from the Euler-Lagrange equation from the Lagrangian $L$ for a free particle, ie.
\begin{equation}
L = T_g = \frac12 g_{ij} (\bm{x}) \dot{x}^i \dot{x}^j.
\end{equation}
We define lagrangians of the mechanical type for systems with configuration space $M$, $L \in C^\infty (TM)$, by choosing a pseudo-Riemannian structure $g$ on $M$ and a potential function $V \in C^\infty (TM)$ as follows
$$L (\bm{x}, \dot{\bm{x}}) = \frac12 g_x (\dot{\bm{x}},\dot{\bm{x}}) - V(\bm{x}) = \frac12 g_{ij} (\bm{x}) \dot{x}^i \dot{x}^j  - V(\bm{x}).$$
The key concept of the Eisenhart lift is to introduce a new degree of freedom with a new 
co-ordinate, thus replacing configuration space $M$ with $\mathbb{R} \times M$. Eisnehart demonstrated the 
possibility of relating the dynamical trajectories of a lagrangian mechanical system with a 
projection on $M$ of etremal length curves on an extended manifold $\widetilde{M} = \mathbb{R} \times M$ with the 
Riemannian structure
$$\widetilde{g} = \Pi_2^* g - \frac1{2V} dz \otimes dz.$$
where
$$\Pi_{1,2}: \mathbb{R} \times M \longrightarrow \mathbb{R}, M.$$
If we assume $g_{00}$ as a function $A$ of the co-ordinates $(x^1, . . . x^n)$, the square of arc length 
geometry
$$ds^2 = g_{ij} (\bm{x}) dx^i dx^j + A (\bm{x}) dz^2.$$
with the associated motion geometry
\begin{equation}
\label{fmgeom} T_g = \frac12 \Big( g_{ij} (\bm{x}) \dot{x}^i \dot{x}^j + A (\bm{x}) \dot{z}^2 \Big).
\end{equation}
then the equations of motion in terms of arc-length $s$ is given by
$${x^i}'' + \Gamma^i_{jk} {x^j}' {x^k}' - g^{ij} \frac{\partial A}{\partial x^j} (z')^2 = 0.$$
Since $z$ is a cyclical variable, we should have
$$A (\bm{x}) \dot{z} = c \in \mathbb{R}.$$
For each value of the parameter $c$, we can use a new parameter $t = cs$. Then the differential 
equations reduce to
$$\ddot{x}^i + \Gamma^i_{jk} \dot{x}^j \dot{x}^k - g^{ij} \frac1{2 A^2} \frac{\partial A}{\partial x^j} = 0 \qquad \qquad A (\bm{x}) \dot{z} = 1.$$ 
Note that when $c=1$, the parameter $t$ coincides with $s$, and the condition $A (\bm{x}) \dot{z} = 1$ corresponds 
to $p_z = 1$. If we choose $A = (2 V)^{-1}$, then we obtain
\begin{equation}
\ddot{x}^i + \Gamma^i_{jk} \dot{x}^j \dot{x}^k + g^{ij} \frac{\partial V}{\partial x^j} = 0.
\end{equation}
Thus, $\widetilde{g}$ is associated with kinetic energy (\ref{fmgeom}) after Legendre transform leads to the new 
Hamiltonian.
\begin{equation}
H = \frac12 \big( g^{ij} p_i p_j + V p_z^2 \big).
\end{equation}
which coincides with the natural Hamiltonian of mechanical type for $p_z = \sqrt{2}$. One way to understand how it makes a difference is shown in the following subsections.

\numberwithin{equation}{subsection}

\subsection{The Metric without Eisenhart Lift}

We shall first look at the look at the system portrayed orginally without the Eisenhart lift. If 
the given general metric without Eisenhart lift is:
$$d l^2 = h_{ij} ({\bm x}, t) dx^i dx^j + 2 \frac{A_i ({\bm x}, t)}m c dx^i dt - 2 \frac{\Phi ({\bm x}, t)}m \ c^2 dt^2.$$
then the Lagrangian is given by
$$L = \frac m2 h_{ij} ({\bm x}, t) \dot{x}^i \dot{x}^j + A_i ({\bm x}, t) c \dot{x}^i \dot{t} - \Phi ({\bm x}, t) \ c^2 \dot{t}^2.$$
We will have the momenta
$$p_i = \frac{\partial L}{\partial \dot{x}^i} = m h_{ij} ({\bm x}, t) \dot{x}^j + A_i ({\bm x}, t) c \dot{t} \qquad \qquad p_t = \frac{\partial L}{\partial \dot{t}} = A_i ({\bm x}, t) c \dot{x}^i - 2 \Phi ({\bm x}, t) c^2 \dot{t} = - H.$$
The Maupertuis form of the action gives the Lagrangian along the geodesic as:
\begin{equation}
\label{geolag1} L_{geod} = p_\mu \dot{x}^\mu = \frac{\partial L_{geod}}{\partial \dot{x}^\mu} \dot{x}^\mu = p_i \dot{x}^i - H \dot{t}.
\end{equation}
Thus, we will have at least one conserved quantity which is the overall Legendre Hamiltonian:
\[ \begin{split}
\frac{d L_{geod}}{d \tau} &= \frac{\partial L_{geod}}{\partial x^\mu} \dot{x}^\mu + \frac{\partial L_{geod}}{\partial \dot{x}^\mu} \ddot{x}^\mu = \underbrace{\bigg[ \frac{\partial L_{geod}}{\partial x^\mu} - \frac{d \ }{d \tau} \bigg(\frac{\partial L_{geod}}{\partial \dot{x}^\mu} \bigg) \bigg]}_0 \dot{x}^\mu + \frac{d \ }{d \tau} \bigg(\frac{\partial L_{geod}}{\partial \dot{x}^\mu} \dot{x}^\mu \bigg), \\
\Rightarrow \qquad &\frac{d \ }{d \tau} \bigg(\frac{\partial L_{geod}}{\partial \dot{x}^\mu} \dot{x}^\mu - L_{geod} \bigg) = 0 \qquad \Rightarrow \qquad \mathcal{H} = \frac{\partial L_{geod}}{\partial \dot{x}^\mu} \dot{x}^\mu - L_{geod} = 0 = conserved.
\end{split} \]
Now, depending on the metric's dependence on time, we will face different situations.

\subsubsection*{Time-Independent Case}

When independent of time $t$, we will have another conserved quantity $H$ in addition to $\mathcal{H}$
$$- H = \frac{\partial L}{\partial \dot{t}} = conserved.$$
From (\ref{geolag1}), we can see that this conserved quantity under time parametrization ($\dot{t} = 1$) is 
given by
$$H = p_i \dot{x}^i - L_{geod} = \frac1{2m} h^{ij} ({\bm x}) (p_i - c A_i) (p_j - c A_j) + \Phi ({\bm x}).$$
Thus, we have 2 conserved quantities: $\mathcal{H}$ and $H$.

\subsubsection*{Time-Dependent Case}

If the metric is time-dependent, $H$ will not be a conserved quantity. This means that we are 
forced to resort to $\mathcal{H}$ as the only conserved quantity.

\subsection{The Metric with Eisenhart Lift}

This time, we will modify the metric with the Eisenhart lift by introducing a dummy variable 
$\sigma$. If the given general metric with Eisenhart lift is:
$$d l^2 = h_{ij} ({\bm x}, t) dx^i dx^j + 2 c \ dt \ d \sigma + 2 \frac{A_i ({\bm x}, t)}m dx^i dt - \frac{2 \Phi ({\bm x}, t)}m \ c^2 dt^2.$$
where the metric is independent of $\sigma$, then the Lagrangian is given by
\begin{equation}
\label{edlag} L = \frac m2 h_{ij} ({\bm x}, t) \dot{x}^i \dot{x}^j + m c \dot{t} \dot{\sigma} + A_i ({\bm x}, t) c \dot{x}^i \dot{t} - \Phi ({\bm x}, t) \ c^2 \dot{t}^2.
\end{equation}
We will have the momenta, where one is a conserved quantity
$$p_i = \frac{\partial L}{\partial \dot{x}^i} \qquad p_t = \frac{\partial L}{\partial \dot{t}} = m c \dot{\sigma} + A_i ({\bm x}, t) c \dot{x}^i - 2 \Phi ({\bm x}, t) c^2 \dot{t} \qquad p_\sigma = \frac{\partial L}{\partial \dot{\sigma}} = m c \dot{t} = conserved.$$
The Maupertuis form of the action gives the Lagrangian along the geodesic as:
\begin{equation}
\label{geolag2} L_{geod} = p_\mu \dot{x}^\mu = \frac{\partial L_{geod}}{\partial \dot{x}^\mu} \dot{x}^\mu = p_i \dot{x}^i + p_t \dot{t} + p_\sigma \dot{\sigma} = p_i \dot{x}^i + p_\sigma \dot{\sigma} + \frac{p_t p_\sigma}{mc}.
\end{equation}
As before, we will have the overall Legendre Hamiltonian $\mathcal{H}$ as a conserved quantity. Now we 
look at the cases of the metric's dependence on time.

\subsubsection*{Time-Independent Case}

When independent of time $t$, as before we have another conserved quantity $p_t$. From (\ref{geolag2}), we 
can see that this conserved quantity is given by
$$- \frac{p_\sigma p_t}{mc} = p_i \dot{x}^i + p_\sigma \dot{\sigma} - L_{geod} = \frac1{2m} h^{ij} ({\bm x}) \bigg( p_i - \frac{p_\sigma}m A_i \bigg) \bigg(p_j - \frac{p_\sigma}m A_j \bigg) + \Phi ({\bm x}) \bigg( \frac{p_\sigma}m \bigg)^2 = H.$$
Thus, we have 3 conserved quantities: $\mathcal{H}$, $H$ and $p_\sigma$.

\subsubsection*{Time-Dependent Case}

If the metric is time-dependent, $H$ will not be a conserved quantity. This means that
$$\mathcal{H} = \big( p_i \dot{x}^i + p_\sigma \dot{\sigma} - L_{geod} \big) + p_t \dot{t} = H + p_t \dot{t} = mc H + p_t p_\sigma = 0,$$
\begin{equation}
\label{eisensur} \therefore \quad p_\sigma = - \frac{m c H}{p_t} = conserved.
\end{equation}
Thus, we have 2 conserved quantities: $\mathcal{H}$ and $p_\sigma$.  \\ \\
Thus, we can say that the Eisenhart-Duval lift is a useful tool for dealing with time-dependent 
systems by giving another conserved quantity $p_\sigma$ to replace the natural Hamiltonian $H$ normally 
used to parametrize motion on the cotangent space.

\subsection{Formulation}

In this section, we will demonstrate the deduction of the Jacobi-metric for time-dependent 
systems. The formulation has been deduced only with the metric line element. \\ \\
Consider the following spacetime metric:
\begin{equation}
d l^2 = c^2 V^2 ( \bm{x}, t ) d t^2 + 2c \ d \sigma \ d t - g_{ij} ( \bm{x}, t ) dx^i dx^j.
\end{equation}
Its corresponding line-element lagrangian is given as:
\begin{equation}
L ( \bm{x}, \dot{\bm{x}}, t ) = m \sqrt{c^2 V^2 ( \bm{x}, t ) \dot{t}^2 + 2c \dot{\sigma} \dot{t} - g_{ij} ( \bm{x}, t ) \dot{x}^i \dot{x}^j}.
\end{equation}
and the momentum conjugate to co-ordinates are given by:
\begin{equation}
\begin{split}
\frac{p_t}c = \frac{\partial L}{\partial \dot{t}} &= \frac{m \big[ c^2 V^2 ( \bm{x}, t ) \dot{t} + c \dot{\sigma} \big]}{\sqrt{c^2 V^2 ( \bm{x}, t ) \dot{t}^2 + 2c \dot{\sigma} \dot{t} - g_{ij} ( \bm{x}, t ) \dot{x}^i \dot{x}^j}}, \\
\frac{p_i}c = \frac{\partial L}{\partial \dot{x}^i} &= \frac{- m g_{ij} ( \bm{x}, t ) \dot{x}^j}{\sqrt{c^2 V^2 ( \bm{x}, t ) \dot{t}^2 + 2c \dot{\sigma} \dot{t} - g_{ij} ( \bm{x}, t ) \dot{x}^i \dot{x}^j}}, \\
\frac{p_\sigma}c = \frac{\partial L}{\partial \dot{\sigma}} &= \frac{mc \dot{t}}{\sqrt{c^2 V^2 ( \bm{x}, t ) \dot{t}^2 + 2c \dot{\sigma} \dot{t} - g_{ij} ( \bm{x}, t ) \dot{x}^i \dot{x}^j}} = q.
\end{split}
\end{equation}
Using the equation for the Maupertuis form of the action
$$L_{geod} = p_i \dot{x}^i + p_t \dot{t} + p_\sigma \dot{\sigma}.$$
we can deduce that for the line element, the relativistic energy equation is:
\begin{equation}
\label{tdmaup} 2c p_t p_\sigma = c^2 g^{ij} p_i p_j + c^2 V^2 ( \bm{x}, t ) p_\sigma^2 + m^2 c^4 = Q^2.
\end{equation}
With the following calculations we find that
\[ \begin{split}
g^{ij} (\bm{x}, t) p_i p_j &= \frac{m^2 c^2 g_{ij} ( \bm{x}, t ) \dot{x}^i \dot{x}^j}{c^2 V^2 ( \bm{x}, t ) \dot{t}^2 + 2c \dot{\sigma} \dot{t} - g_{ij} ( \bm{x}, t ) \dot{x}^i \dot{x}^j}, \\
\Rightarrow \qquad m^2 c^2 + g^{ij} (\bm{x}, t) p_i p_j &= \frac{m^2 c^2 \big( c^2 V^2 ( \bm{x}, t ) \dot{t}^2 + 2c \dot{\sigma} \dot{t} \big)}{c^2 V^2 ( \bm{x}, t ) \dot{t}^2 + 2c \dot{\sigma} \dot{t} - g_{ij} ( \bm{x}, t ) \dot{x}^i \dot{x}^j} = 2q p_t - q^2 c^2 V^2 ( \bm{x}, t ), \\
\Rightarrow \qquad &\frac{g^{ij} (\bm{x}, t)}{2 q p_t - q^2 c^2 V^2 ( \bm{x}, t ) - m^2 c^2} p_i p_j = 1.
\end{split} \]
this result can be written by writing $V^2 ( \bm{x}, t ) = 2m U ( \bm{x}, t )$ as:
$$\frac{c^2 g^{ij} (\bm{x}, t)}{2 \big[ c^2 q p_t - q^2 c^4 U ( \bm{x}, t ) \big] - m^2 c^4} p_i p_j = 1.$$
Thus the time-dependent Jacobi-metric is given by:
\begin{equation}
\label{tindepjmet}
\begin{split} 
J^{ij} ( \bm{x}, t ) &= \frac{g^{ij} (\bm{x}, t)}{2 \big[ q p_t - q^2 U ( \bm{x}, t ) \big] - m^2 c^2}, \\
J_{ij} ( \bm{x}, t ) &= \big[ 2 \big\{ q p_t - q^2 U ( \bm{x}, t ) \big\} - m^2 c^2 \big] g_{ij} (\bm{x}, t).
\end{split}
\end{equation}
which projects the geodesic onto the constant momentum hypersurface $\dfrac{p_\sigma}c = q$. If we employ 
the following approximation:
$$Q = mc^2 + \mathcal{E} (t) \quad \Rightarrow \quad \bigg( \frac Q{mc^2} \bigg)^2 \approx 1 + \frac{2 \mathcal{E} (t)}{mc^2} \qquad (\mathcal{E} << mc^2),$$
\begin{equation}
\label{tdapp} \therefore \qquad 2 q p_t \approx m^2 c^2 + 2 m \mathcal{E} (t).
\end{equation}
then the Jacobi metric under (\ref{tdapp}) will approximate to:
\begin{equation}
\label{tdepjmet} J_{ij} ( \bm{x}, t ) = 2m \big[ \mathcal{E} (t) - q^2 U ( \bm{x}, t ) \big] g_{ij} (\bm{x}, t).
\end{equation}
which is the non-relativistic approximation for time-dependent systems modified by an Eisenhart-
Duval lift. One may attempt to verify this by deducing the formulation starting 
from the mechanical Lagrangian (\ref{edlag}), or use a projective transformation as shown in the next section.

\numberwithin{equation}{section}

\section{Comparison to Projective Transformation}

Projective geometry can be used to describe natural Hamiltonian systems and generate the dualities 
between them. The Jacobi metric can be alternatively formulated from a projective transformation 
in the phase space as described in \cite{mcar}. This is described by the null Hamiltonian, for 
which the curve is parametrised by the arc length.
\begin{equation}
\label{nullham} \mathcal{H} = \frac1{2m} g^{ij} (\bm{x}) p_i p_j + U (\bm{x}) p_u^2 - \text{sgn}(H) p_y^2.
\end{equation}
Upon setting $p_u^2 = 1$ and $p_y^2 = |E|$, where $E$ is the energy, we get null geodesics that project 
down to the original system. Rescaling the Hamiltonian by the factor $\Omega^2 = E - V(\bm{x})$, gives
$$\widetilde{\mathcal{H}} = \frac{\mathcal{H}}{\Omega^2} = \frac1{2m} \frac{g^{ij} (\bm{x}) p_i p_j}{E - U (\bm{x})} - 1.$$
This is the null geodesic Hamiltonian related to the Jacobi metric for time-independent systems. 
From it, the inverse metric, we can deduce the Jacobi metric (\ref{jacmet}). 
\begin{equation}
\label{proj1} \widetilde{g}^{ij} (\bm{x}) = \frac1{2m} \frac{g^{ij} (\bm{x})}{E - U (\bm{x})} \qquad \qquad \widetilde{g}_{ij} (\bm{x}) = 2m \big[ E - U (\bm{x}) \big] g_{ij} (\bm{x}).
\end{equation}
To account for time-dependence, as previously we modify the null Hamiltonian (\ref{nullham}) via an Eisenhart-Duval such that $- \text{sgn}(H) p_y^2 \ \longrightarrow \ \frac{p_u p_v}{mc}$ 
to include the extra co-ordinate as a dummy variable
\begin{equation}
\label{nulleduval} \mathcal{H} = \frac1{2m} g^{ij} (\bm{x}, t) p_i p_j + U (\bm{x}, t) p_u^2 + \frac{p_u p_v}{mc}.
\end{equation}
which is essentially an Eisenhart-Duval lifted Hamiltonian. As before, if we rescale (\ref{nulleduval}) by 
a factor $\Omega^2 = - p_u p_v - U(\bm{x}, t) p_u^2$, then we will get the corresponding null geodesic Hamiltonian 
for the Jacobi metric.
$$\widetilde{\mathcal{H}} = \frac{\mathcal{H}}{\Omega^2} = - \frac1{2m} \frac{g^{ij} (\bm{x}, t) p_i p_j}{p_u p_v + U (\bm{x}, t) p_u^2} - 1.$$
Here, if we write $p_u = q = mc$, and $p_v = - \mathcal{E} (t)$ in accordance with (\ref{eisensur}), we will have the 
Jacobi metric for time-dependent systems 
\begin{equation}
\label{proj2} \widetilde{g}^{ij} (\bm{x}, t) =  \frac1{2m} \frac{g^{ij} (\bm{x}, t)}{\mathcal{E} (t) - q^2 U (\bm{x}, t)} \qquad \qquad \widetilde{g}_{ij} (\bm{x}, t) =  2m \big[\mathcal{E} (t) - q^2 U (\bm{x}, t)\big] g_{ij} (\bm{x}, t).
\end{equation}
Upon comparison (\ref{proj1}) and (\ref{proj2}) match (\ref{jacmet}) and (\ref{tdepjmet}) respectively. This shows that the 
Jacobi metric in the non-relativistic limit can be deduced from projective transformations of 
time-dependent systems, just as \cite{mcar} demonstrates it for time-independent systems.

\section{Conclusion and Discussion}

The Jacobi metric formulated by one of us was shown to derive from the metric line element, 
preserving the angular momentum as a conserved quantity and acting as a conformally flat 
metric for cases like Kepler or $n$-body problems. This seems to be distinct from the Jacobi-
metric formulated from the Lagrangian and Hamiltonian in the usual mechanical formulation, 
projecting a geodesic in $n+1$ spacetime into dynamics in $n$ dimensions. However, applying non-
relativistic approximations to the line element formulation shows that the two approaches are 
equivalent. With this in mind, we deduced the Jacobi metric in relativistic and non-relativistic 
form for various metrics: Taub-NUT space, Bertrand spacetime and the Kerr spacetime. \\

So far, the Jacobi-metric has been formulated only for autonomous systems. This was 
possible because of a conserved quantity made available via a cyclical co-ordinate. For autonomous 
systems, the Hamiltonian is the relevant conserved quantity, conjugate to time as the cyclical 
variable. However, such convenience is denied in the case of time-dependent systems. Under 
such circumstances, the Eisenhart-Duval lift proved to be a useful tool, by providing a dummy 
variable along an extra dimension, and thus, a conserved quantity. This gives us a momentum 
equation from which we can define a metric for the unit momentum sphere, and thus, the 
Jacobi metric for time-dependent systems.

\section*{Acknowledgement}

We thank Marco Cariglia for his correspondence and contribution to this project, and referring us to his article 
\cite{mcar}, which were invauable in developing this article.

\end{document}